\RequirePackage{fix-cm}
\documentclass{article}
\usepackage[margin=22mm]{geometry}
\usepackage{setspace}
\usepackage{graphicx}
\usepackage{caption}
\usepackage[square, comma, sort&compress]{natbib}
\usepackage{amssymb}
\usepackage{algorithm}
\usepackage{algorithmic}
\usepackage{amsmath}
\usepackage{multirow}

\begin{document}

\title{Internet Control Plane Event Identification using Model Based Change Point Detection Techniques %\thanks{Grants or other notes
}
%\subtitle{Control Plane Event Identification in Internet}


%\author{S.P.Meenakshi         \and
%        S.V.Raghavan %etc.
%}
\author{S.P.Meenakshi\\ Department of Computer Science and Engineering \\ IIT Madras, India \\ spmeena@cse.iitm.ac.in        \and
        S.V.Raghavan \\ Department of Computer Science and Engineering \\ IIT Madras, India \\ svr@cse.iitm.ac.in%etc.
}

\date{\vspace{-5ex}}

%\institute{Indian Institute of Technology Madras \at
%              Chennai -600036, India. \\
%              Tel.: +914422575374\\
%              \email{spmeena@cse.iitm.ac.in}           %  \\
%            Indian Institute of Technology Madras \at
%            Chennai -600036, India \\
%            Tel.: +914422575374\\
%            \email{svr@cse.iitm.ac.in}
%}


\maketitle

\begin{abstract}
In the raise of many global organizations deploying their data centers and content services in India, 
the prefix reachability performance study from global destinations garners our attention.
 The events such as link/node failures and DDoS attacks occurring in the Internet topology have
 impact on Autonomous System (AS) paths announced in the control plane and reachability of prefixes from spatially distributed ASes.
 As a consequence the customer reachability  to the services in terms of increased latency and outages for a short 
or long time are occurring.
The challenge in control plane event detection is when the data plane traffic is able to reach the intended destinations
correctly.  However detection of such events are  crucial for the operations of content and data center industries.
To capture the events we need extreme monitoring infrastructure. 
By monitoring the spatially distributed routing table features like AS path length distributions, 
spatial prefix reachability distribution 
and covering to overlap route ratio, we can detect the control plane events. In our work, we study prefix AS 
paths from
  the publicly available route-view data and analyze the global reachability as well as reachability to Indian
 AS topology. The  temporal
pattern analysis on distributed peer prefix announcements are performed to understand the occurrence of Internet
wide events. To capture the spatial events in a single temporal pattern, we propose a  counting based measure
using prefixes announced by \textbf{\emph{x}} $\%$ of spatial peers.   
We measure and characterize prefix reachability by using this measure.
Employing statistical characteristics change point detection and temporal aberration algorithm on the time series of 
the proposed measure, we identify the occurrence of long and stochastic control plane events. 
The impact and duration of the events are also quantified.
We validate the mechanisms over the proposed measure using the SEA-Me-We4 cable cut event manifestations 
in the control plane of Indian AS topology. The cable cut events occurred on 6th June  2012 (long term event)
and 17th April 2012 (stochastic event) are considered.  Other evidences like upstream changes
of major transit and originating ASes of Indian AS topology and peering changes in the route-view table are correlated to
confirm the occurrence of the control plane events.
\end{abstract}
\doublespacing
\section{Introduction}
\label{intro}
In the raise of many global organizations deploying their data centers and content
 services in India,
 the prefix reachability performance study from global destinations garners our attention. Particularly
online services provided from Indian companies to its European and US clients heavily rely on the 
Internet. The outages due to cable cuts and its impact on bandwidth and delay to the Indian industries
are reported in news reports \citep{SMW42008Jan,SMW42008Dec,SMW42012Jun}. 
The events such as link/node failures and DDoS attacks occurring in the Internet topology have
 impact on Autonomous System (AS) paths and reachability of prefixes from spatially distributed ASes.
As a consequence the customer reachability  to the services in terms of increased latency and outages for a short 
or long time are occurring.
 The events occurring in the control plane can be classified as  stochastic and long term events. 
Stochastic events are defined as the events that have performance impact of one day. 
Long term events are defined as the events that have performance impact that prolongs  more than a day.
The performance impact can be quantified using prefix reachability count and the duration.\\ 
\indent
When the data plane traffic is not  able to reach the intended destinations, the outages
are perceived immediately and reported in the blogs and news reports. In case of the data plane traffic
 is being able to reach intended destinations correctly with degraded performance then the events 
happening in the control plane are not getting much attention. 
But identifying these events are crucial for the operations
 of content, data and call center
 industries. The Sea-Me-We4 submarine cable that interconnects India to Europe and Middle East got
cut between Malaysia and Thailand on June 2012 \citep{SMW42012Jun}. But the traffic is rerouted
via other routes with increased latency.  To capture such events we need a distributed monitoring infrastructure.
 By monitoring the features like AS path length distributions, spatial prefix reachability distributions
 and covering to overlap route ratio we can detect the control plane events.\\
\indent
 The publicly available route-view data provides \citep{routeview} the control plane data
collected from its geographically  distributed peers.
  Using this data we can characterize the required features for the problem under consideration.
We have analyzed the temporal prefix reachability patterns from spatially located peers
with respect to Indian AS topology. Based on the observations, we propose a measure that captures
the Internet wide control plane activities in a single temporal pattern. The measure counts the
reachable prefixes to a country level AS topology from spatially distributed locations. Here the
locations are inferred from the locations of peers. The temporal pattern constructed using the measure is 
analyzed for time series modeling.\\
\indent
 In our work, time series model is  used to capture the long and short term prefix announcement deviant behaviors
that are manifestations of events. ARIMA models are used for generating the time series with different characteristics.
To make the non stationary time series compliant for ARIMA modeling, mean change point detection
technique is used to identify stationary segments in the time series. Temporal aberration algorithm
is used on ARIMA forecasting to capture the stochastic events. Long term events are identified
using the change point detection technique. We validate the mechanisms applied on the proposed measure
  using the SEA-Me-We4  cable cut events that have performance  impact on Indian AS topology.\\
\indent
The paper is organized as follows. BGP anomalous behavior study carried out in the literature 
for different large scale events are discussed in section 2. Route-view data and country level analysis for
Indian AS topology are discussed in section 3. Section 4 deals with measure selection criteria. 
 The model based event identification mechanisms are presented in section 5. The mechanisms 
validation is done in section 6 followed by conclusion in section 7.
\section{Related Work}
\label{sec:1}
In literature on BGP deviant behavior analysis, the measure widely used to identify the abnormal behavior 
is announce and withdraw reachability information available in BGP updates. The  studies on BGP instabilities 
and dynamics \citep{routeInstability1,routedDynamics1} consider  10 distinct BGP attributes from update messages
to summarize every minute BGP activity. The current and a decade ago BGP dynamics are compared using
 the BGP attributes in the study. It is found that the forwarding dynamics are dominant and also
 consistent across different days.  
The BGP activities exhibit major changes in the number of announced or withdrawn prefixes from the ASes
 due to the occurence of unusual events. The events such as  prefix interception attack \citep{prefixInterception1},
 prefix hijacking attacks \citep{prefixhijack1}, political censorships \citep{politicalCensor1,politicalCensor2},
 hurricanes, and cable cuts \citep{iSeismograph} are studied using BGP data as the primary data source.
  The BGP update attributes are used in \citep{iSeismograph} to measure the impact of Internet earthquakes
by analyzing the deviant behavior of the attributes. 
The BGP routes from the
 updates to the respective countries \citep{politicalCensor1,politicalCensor2} and Youtube query 
traffic are measured to identify the
 Internet blackouts. The routing paths to geographically distributed web servers are \citep{firewallCensor1}
 used to understand 
the AS and router level firewall infrastructure for keyword based filtering. Average prefix path length and 
trace route latency measurements are used to identify the detoured Facebook traffic via China telecom
 \citep{prefixInterception1}. The root cause of the event is identified as reduced prepended AS path announcement
to Facebook prefixes.
 Prefix hijacking detection approaches use inconsistencies in route advertisements, route qualities and
hop count monitoring from deployed vantage points.
 In our work, we use  time series of  prefix reachability 
count  to a country level AS topology announced by x $\%$ spatially distributed peers to identify the events.
The measure is used to quantify the reachability impact and duration of long and short term events.
\\
\indent 
The events are captured based on the deviant behavior from the normal activity. Different mechanisms are used
 to identify the abnormal behavior. The work on I-seismograph \citep{iSeismograph} uses various 
BGP attributes to identify the deviant behavior. The K-Medoids clustering algorithm is used to 
cluster the attributes into normal and abnormal states. The impact  of a single data bin 
 is computed with respect to the normal cluster.  The impact during the monitoring period is computed based on how 
all the data bins collectively deviate from the normal cluster. In our work, we use mean change point detection
 algorithm  on time series  of prefix reachability count announced by spatially distributed peers, for 
detecting long term events. A mean change in the segment using identified threshold i.e., greater than 15 $\%$ 
of total prefix count, 
is considered as a significant long term event. Temporal aberration algorithm is used to detect
 stochastic events. The impact of both the events are quantified using the prefix reachability
count with respect to a moving average. The time series modeling  of the measure has an advantage that it 
can be correlated  with other time series measures such as major link events and attacks to infer the 
root cause for the occurrence of the events.
\section{Route-view Data and Country Level  Analysis}
\begin{figure}
 \centering
 \includegraphics[width=0.5\textwidth]{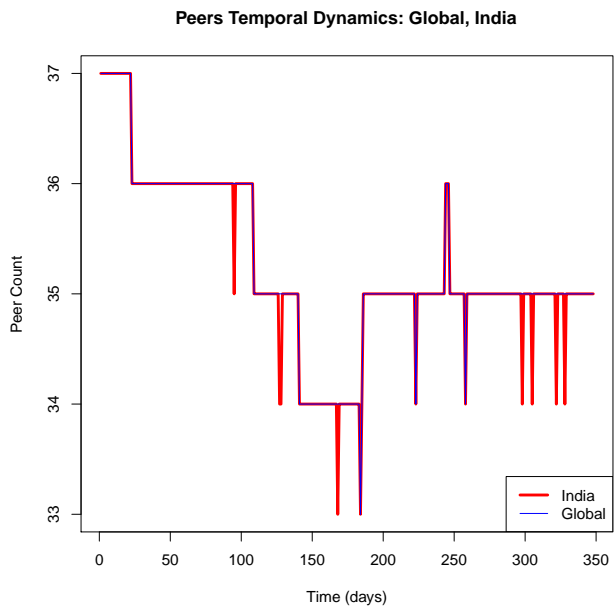}
 \caption{Peers Temporal Dynamics}
 \label{fig:peerdynamics}
\end{figure}

\begin{figure}
 \centering
 \includegraphics[width=0.5\textwidth]{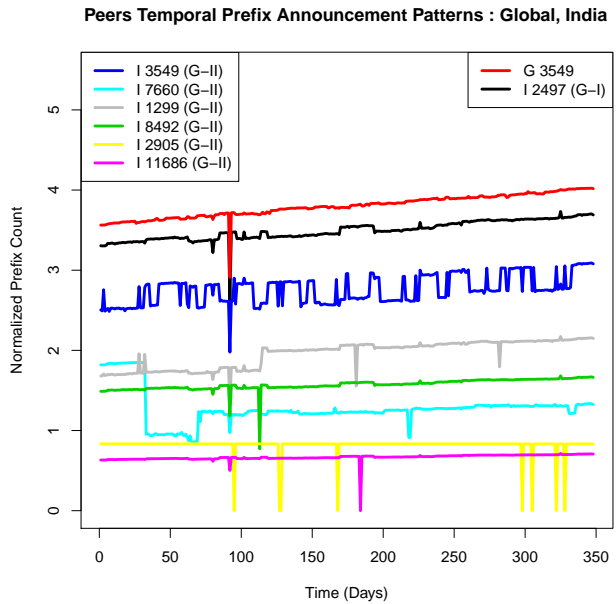}
 \caption{Prefix Announcement Temporal Dynamics From Different Peers}
 \label{fig:prefixtemporal}
\end{figure}

Route-view snapshot carries route information collected from a maximum of 38 peers located in different places. The BGP
feeds collected with an interval of two hours are archived as time stamped snapshots in \citep{routeview}.
The mandatory BGP route attributes are prefixes, next-hop and AS path. The prefixes
are entered in ascending order of their numerical values along with other attributes. 
The next-hop attribute of the BGP route
provides the information on advertising ASes IP address. In route entries of the snapshot, the next-hop indicates the
 IP address of the peer. Two different next-hop address to an AS peer can be interpreted as the BGP feeds
are collected from two BGP border routers of the same AS.
The AS path entries for each prefix is a sequence of ASes starting with the peer AS, followed by intermediate ASes
in the path and end with the AS that originates the prefix. 
 The AS paths announced for a prefix
by different peers are entered consecutively in the snapshot.  
 Hence counting a prefix consecutively until 
a change in prefix occurs provides the information 
 from how many peers a prefix can be reachable at that time. 
 We have extracted and grouped the geographically
 located peers from the snapshot based on  continents. This exercise is to understand the number of 
representative peers in each continent. The distributions of 
the peer locations in five continents are given in 
the table \ref{table:peerdist}. The continents Asia, Australia and Africa have less than 10 percentage of total peers.
South America has no peer representation at all.\\
\begin{table*}
\caption{Peer Spatial Location Distribution}
\label{table:peerdist}       % Give a unique label
\begin{center}
 
\begin{tabular}{|c|c|c|c|c|}
\hline\noalign{\smallskip}
 Asia & Europe & North America & Australia & Africa \\
\noalign{\smallskip}\hline\noalign{\smallskip}
\hline

2 & 10 &  23 &  1 & 2 \\ 
\noalign{\smallskip}\hline
 \end{tabular}
\end{center}
\end{table*}
\indent
We have considered daily snapshots that are time stamped to 0000 hours from 01-01-2012 to 31-12-2012 in our work.
 Using APNIC \citep{APNIC}
 Regional Internet Registry (RIR), we extracted allocated Indian AS numbers. The APNIC RIR contains the static record
of AS numbers, allocated IPv4 and IPv6 prefixes for countries in Asia Pacific Region. The extracted Indian AS numbers
 are used to filter the routes for Indian ASes from the daily snapshots. The  prefix counts announced 
globally and to Indian ASes by different peers are given in table \ref{table:PrefixCntsGI}. Hereafter we refer
 the path announcements to all the prefixes in complete snapshot as global.
\begin{table*}
\caption{Announced Prefix Counts: Global and India }
\label{table:PrefixCntsGI}       % Give a unique label
\begin{center}
 
\begin{tabular}{|l|c|c|p{5cm}|p{5cm}|}
\hline\noalign{\smallskip}
 Date & Type & Peer Count & Peers & Prefix Counts \\
\noalign{\smallskip}\hline\noalign{\smallskip}
\hline
\multirow{2}{*}{01-01-2012} & Global & 37 & 286 293 701 812 852 852 1221 1239 1299 1668 2152 2497 2905 2914 3130 3130 3257 3303 3356 3549 3549 3561 5056 5413 6539 6762 6939 7018 7660 8001 8492 11537 11686 13030 22388 31500 39756 & 380992 384469 379432 373237 381070 381067 381843 379792 376669 379936 382751 381453 2474 380729 381143 381135 380480 150422 378602 380738 378630 379754 381341 380096 379826 381433 382521 379882 384557 364665 385139 13435 388438 382100 13575 389785 382585 \\
\cline{2-5}
& India& 37 & 286 293 701 812 852 852 1221 1239 1299 1668 2152 2497 2905 2914 3130 3130 3257 3303 3356 3549 3549 3561 5056 5413 6539 6762 6939 7018 7660 8001 8492 11537 11686 13030 22388 31500 39756 & 17135 17160 17136 17139 17132 17132 17150 17133 14915 17135 17141 17153 1 17134 17150 17150 17148 4677 17132 15565 17134 17133 17103 17115 17116 17134 17134 17134 17178 15256 17158 289 17159 17149 289 17472 17148\\
\hline
 \end{tabular}
\end{center}
\end{table*}
We have observed in the route-view snapshots that peers with an average of one, dropped or included with varying time
 intervals. For instance, the peer AS located in Romania : AS 39756 ceased the peering from 27-01-2012 to 
28-08-2012 and resumed on 29-08-2012. The peer AS 3741 located in South Africa started the peering only from
  22-07-2012.
Peers temporal dynamics with respect to global and Indian AS topology is 
given in figure \ref{fig:peerdynamics}. 
  From the figure \ref{fig:peerdynamics}, we can see that there is an occasional peer count drop of one to Indian AS 
topology  but not in the global level. After analyzing the extracted data, the peer involved in such drop is 
identified as AS 2905 located in South Africa. It announces only one prefix that belongs to Google corporate network
AS 45566 and occasionally drops announcing the prefix.\\  
\indent
Next, we have analyzed the temporal prefix announcement patterns extracted from each of the 38 peers for India and 
compared it with 
the global level patterns. In the normalized scale, 25 of the global level temporal patterns exhibit a linear trend with only one
significant drop event that has occurred on 17th April 2012. For normalization, we use a scale down factor computed 
as difference between maximum and minimum temporal values. The extracted prefix announcement patterns to India
 significantly 
varies in 18 out of 38 peers by exhibiting different raise and drop events. The rest of the 20 peers
 prefix announcements  have similar
 temporal patterns with three visibly significant events. First one is a long term drop event from 15th March to 
5th April.
Second one is a stochastic drop event in prefix announcement that
appears on 17th April. Third one is long term raise event that appears from 7th July  to 29th July.  
 The second and third  events co-inside with the time of cable cut occurred in  SEA-Me-We4 submarine cable.
For the purpose of further analysis, we group the peers into two groups based on the aforesaid observations.
 Peer ASes with
similar temporal prefix announcement patterns (20 peers) are placed in group I. Group II is placed with
 peers having different prefix 
announcement patterns (18 peers). 
The peer ASes and their locations in Group I and Group II are given in table \ref{table:peergroup}.
\begin{table*}
\caption{Peer Groups and Locations }
\label{table:peergroup}       % Give a unique label
\begin{center}
 
\begin{tabular}{|c|p{10cm}|}
\hline\noalign{\smallskip}
  Type & Peers and Locations \\
\noalign{\smallskip}\hline\noalign{\smallskip}
\hline
Group I  & 1221 (Australia) ,2497 (Japan), 293, 701, 1239,1668, 2152, 2914, 3130, 3356, 5056, 6939,7018 (USA),
 852, 6539 (Canada) 286 (Netherlands),3257 (Germany),5413 (UK) ,6762 (Italy) \\
\hline
Group II & 7660 (Japan),3549, 8001, 11537, 11686, 22388 (USA), 812 (Canada),1299 (Sweden), 8492,31500 (Russia),
3303,13030 (Switzerland),39756 (Romania), 2905,3741 (South Africa)  \\
\hline
 \end{tabular}
\end{center}
\end{table*}
The representative temporal prefix announcement patterns for global category, Group I and Group II categories for India
 are given in figure \ref{fig:prefixtemporal}.
The varying temporal patterns from different peers invoke the intuition that the stochastic and long term prefix 
drop/raise events occur network wide
 with different magnitudes (prefix counts) due to various causes. At a time point, the events are manifested in one or
multiple peer temporal patterns. But in a countrywide view, we are interested to find answers for the following
questions. How many prefixes are reachable 
from maximum number of peer locations?
 How the prefix announcements from maximum number of peers vary temporally? These two questions will
answer the impact on prefix reachability from different locations and the duration of the event.
 To address these two questions we are proposing a spatio temporal prefix counting measure. 
 \section{Measure Selection Criteria}
\label{sec:3}
Spatio temporal measures provide  good indication for control plane events that take place spatially
in the Internet. The measures are combination of two components namely long term trend and  irregular
variations. Long term trend is induced by raise in number of new content and data center
service networks. The irregular variations are due to node/link failures, popular events and DDoS attacks.
We are interested to identify the irregular variations and find out the specific times the impact of the events
is higher or lower. The prefix count reachability from different peer locations is the considered feature 
measure in our work. The spatial locations are interpreted from the spatially located peers associated
 with BGP feeds to the route-view servers. Noticeable deviations in the number of prefixes reachable from x $\%$ of 
spatially distributed peers often indicate undesirable control plane events. The value for x is chosen based on the
number of maximum prefix announcement peers for a country level. The rationality behind choosing
the percentage of peers and the value for it is discussed in section \ref{sec:3.1}.\\
\indent 
 Our general approach to the
 control plane event detection problem is to first establish a measure that reflects the global Internet
operating conditions (  prefix count reachable from \textbf{\emph{x}} $\%$ of spatially distributed peers)
 whose expected behavior 
we can discern from individual peer temporal pattern analysis and then determine the duration and impact of the deviant 
behavior. Two techniques are employed on the measure for the purpose of identifying long  and short term prefix 
based events.
 By considering statistical characteristics change in the long term events, mean change point detection
technique is used to detect and quantify that kind of events. ARIMA models are used in short term event detection
 and quantification.
Changing network conditions and continuously evolving prefix patterns induce highly fluctuating behavior 
on prefix reachability  measures over time. It is  reasonable to assume that for a limited span of 
time we can model irregular behavior patterns using time series models such as ARIMA.
This model is used to detect one step ahead deviant behavior which is as an indication of stochastic control plane events.
\subsection{The Measure and Empirical Measurement}
\label{sec:3.1}
In the routing table data, we have reachability information for unique prefixes from each of the peers.
There are $X_{t}$ peers, where $x_{1}$, $x_{2}$ \dots  $x_{n}$ are spatially distributed peers at time t. $N_{t}$ is
 maximum number
of peers announcing prefixes at time t. $Y_{t}$ is vector of prefixes $y_{1}$, $y_{2}$ \dots $y_{n}$,
 announced by $X_{t}$. The maximum unique 
prefix value
 $y_{max}$ is computed as $y_{1}$ U $y_{2}$ \dots U $y_{n}$. Since the prefix values are evolving in nature and subjected
to irregular variations, we take average of 7 prior and current values for our analysis at time t. Each of the prefix
is announced by X peers which varies between 1 \dots $N_{t}$ at time t.
 We create different percentage of peer bin ranges and counted the number
of prefixes announced by each peer bin range. The counting measure is interpreted as number prefixes announced by the
percentage of peers at time t. Each bin measurement is compared with $y_{max}$ to find out the relative rank 
at time t. When n peers increase announcing y prefixes each at time t, then the prefix count of one or more peer bin
ranges will increase. Similarly prefix count drop will  occur in one or more peer bin ranges when n peers drop
y prefixes each.
 The temporal variations in the relative
 rank of the peer bins is an indication of  events. This change will also be manifested in the temporal patterns
 of each of the peer bin range. We are interested to identify a peer bin range that announces greater than 95 $\%$ of the prefixes.
Since $y_{i}$  values are not equal and have variations, we want to find out  the peer bin range that announces 
at least 95 $\%$ of $y_{max}$.
 Initially
the peer bins are assigned with the following percentage ranges.
\begin{enumerate}
 \item $>$90 $\%$ 
 \item  81 to 90 $\%$
 \item  51 to 80 $\%$
 \item  28 to 50 $\%$
 \item  14 to 27 $\%$
 \item   6 to 13 $\%$
 \item  $<$ 5 $\%$ 
\end{enumerate}
  The initially assigned bin values can be modified according
to the empirical data under consideration. This procedure is to identify the appropriate bin percentage range that 
announces at least 95 $\%$ of prefixes. The prefix announced in this bin range is considered
temporally to identify the events and the impacts.
\\
\indent
 We have measured the spatio temporal prefix count measure on AS path data to Indian prefixes that are extracted 
from daily route-view snapshots. Further more, time series modeling is employed on the measure for event detection. 
  Each peer feed has unique path entry for
 a prefix in the route-view table. The number of paths are equal to
 number of reachable unique prefixes. We hold this assumption since multihoming contributes to less than 3 $\%$ of
the paths.  The number of paths announced to each prefix  are counted and added to the corresponding peer bin range. 
The initial range of peer bin percentage
 and the prefix count reachable from each  range for Indian AS topology is give in table \ref{table:reachable}
 \begin{table*}
\caption{Prefix Reachability From Range of  Spatial Locations}
\label{table:reachable}       % Give a unique label
\begin{center}
 
\begin{tabular}{|c|c|c|c|c|c|c|c|c|}
\hline\noalign{\smallskip}
 Date & Unique Prefixes& $>$ 90 $\%$ & 81-90 $\%$ & 51-80 $\%$ & 28-50 $\%$ & 14-27$\%$ & 6-13 $\%$ & $<$ 5 $\%$\\
\noalign{\smallskip}\hline\noalign{\smallskip}
\hline

01-01-2012 & 17502 &  13286 & 3813 &  33 & 13 & 47 &  23 & 287 \\ 
01-02-2012 & 17225 &  13256 & 3847 &  16 & 14 & 46 &  28 & 18 \\  
01-03-2012 & 17215 &  13350 & 3744 &  30 & 10 & 46 &  22 & 13 \\  
12-29-2012 & 19294 &  11456 & 7701 &  34 & 16 & 38 &  35 & 14 \\
12-30-2012 & 19292 &  11428 & 7730 &  34 & 14 & 37 &  35 & 14 \\
12-31-2012 & 19222 &  11499 & 7592 &  40 & 8  & 36 &  33 & 14  \\

\noalign{\smallskip}\hline
 \end{tabular}
\end{center}
\end{table*}

From the extracted sample data for the duration of 01-01-2012 to 31-12-2012 given in the table, we find that the 
prefixes grouped under first and second range is  approximately 99 $\%$ of
the unique prefixes. It can be interpreted as 77 $\%$ of the unique prefixes  are announced by 
greater than 90 $\%$ of the peers in different locations. In addition, another 22 $\%$ of the unique prefixes are announced by 80-90 $\%$
range of the peers. Cumulatively 99 $\%$ of the prefixes are announced by more than 80 $\%$ of the peers. In
the observation duration, maximum temporal dynamics occurred only in these two ranges. When there is a drop/increase
 in above 90 $\%$ range, most of the time the equal value of increase/drop is observed in  80-90 $\%$ range. From this, we
can safely infer that less than 10 $\%$  of peers are only involved in temporal dynamics most of the time. 
From the extracted data, we found that route-view routers involved in  peering with approximately 38 ASes. 
In few of the ASes at most two routers are involved in peering activity. 

%In the two peer routers of an AS,
% at time t, only one of the router announce prefix path for india. Peering router dynamics are shown in figure
%\ref{fig:peerdyn}. ------>chk

%\begin{figure}
% \centering
% \includegraphics[width=0.5\textwidth]{./graphs/peer_dynamics.eps}
% \caption{AS Peer Dynamics in Route Announcement to India}
% \label{fig:peerdyn}
%\end{figure}

\begin{figure}
 \centering
 \includegraphics[width=0.5\textwidth]{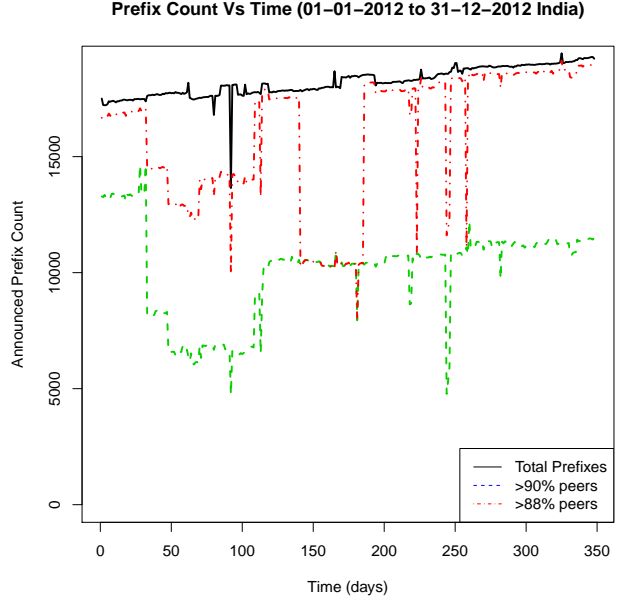}
 \caption{Prefix Vs Time}
 \label{fig:announce}
\end{figure}

Considering the temporal peer dynamics from the aforesaid two ranges, we are interested to identify the maximum percentage of
 peers announcing greater than 95 $\%$ of 
unique prefixes initially. So we computed the prefixes  announced by each percentage of peers  above 80 $\%$
 at time t. The announced prefix count versus percentage of peers announcing the prefixes are given in
 figure \ref{fig:announce}.
When there is an event in the control plane, then it is imperative that the prefix count announced by this spatially
located  peers  will have impact.
We have found that greater than 88 $\%$ of peers announce more than 95 $\%$ of the prefixes.
This peer percentage is more sensitive to control plane events than the above 90 $\%$ peers.  
 Two long time (more than a day) drops 
and five short time drops (one day) that
have impact on more than 10 $\%$ prefixes are observed from the figure \ref{fig:announce}.
Two long time raise events with 1-2 $\%$ increase in prefix count also seen. We compare both greater than 88 and 90 $\%$ peer
temporal patterns  with the temporal patterns of individual peers given in figure \ref{fig:prefixtemporal}.
 It has been
observed that the prefix (drop/raise) events occurred in  individual peer patterns are captured in
 the patterns of both the peer percentages. 
But based on visual interpretation, above 88 $\%$ peer prefix announcements captures more events. 
Hence we take this spatially distributed peer percentage announced temporal prefix counts measure
 for identification of control plane events. The optimal peer percentage that captures all/maximum
events needs further investigation. 
\section{Event Identification}
\label{sec:4}
Events that are happening unusually in the time series data are identified using its deviant behavior from 
time series model forecasts.
For deviant behavior analysis  on time series data, it is necessary that the series should meet a set of ideal conditions,
 such as the data being consistent and trend free. Consistency implies that all the collected data belong to the
 same statistical population (i.e., a generation process with the same parameter  generates all the data). Trend
free implies that the time series should be stationary. The time series data can be described using the data generating
process and the moments such as mean, variance and autocorrelation of order k. Stationarity is an essential
property of time series process. A time series process is said to be covariance-stationary or weakly stationary 
if its first and second moments are time invariant. A stationary process also has mean reverting property. If the
time series holds these properties then only the process model will forecast with good point forecast and 
95$\%$ confidence prediction interval. The accurate forecasting is essential to capture the deviant behaviors in our
case.\\
\indent
 But our time series data is non stationary. We can observe two long duration of drops in our 88 $\%$ peer bin prefix
announcement measure. This indicates parameter change in the generation process. In our measure, non stationarity 
is present due to  factors such as new services, node/link failures
 and attacks. The prefix announcement evolve temporally with a long term trend due to new services. 
Factors such as node/link failures and attacks  have long and stochastic variations on prefix announcements.
The statistical characteristics of the series such as mean changes during long term events. Due to which the time series 
must be generated using the model with different parameter. Hence to obtain the consistency property for the
time series model, we detect the significant mean changes using change point detection technique. Then 
each mean invariant segment is modeled using ARIMA models for forecasting and through which stochastic 
deviant behavior events are captured. \\
\indent
The mean variant characteristics on the identified measure during long term events is used in change point detection
 technique for detecting long term events. Since the spatio temporal measure is dynamic that is subjected to many
 external factors, long time prefix announcement changes are
 reasonable to be expected. Hence mean change point detection can be used to find the significant long
duration prefix  announcement changes.
\\
\indent
Stochastic events are identified using temporal aberration algorithm that uses time series methods.
We use the mean invariant  segments identified by 
the change point detection technique to fit for ARIMA models.
The model parameter values  for each segment is identified using ACF and PACF coefficients of that segment.
Then using threshold value based mechanism on each segment, significant stochastic changes are identified.
\begin{figure}
 \centering
 \includegraphics[width=0.5\textwidth]{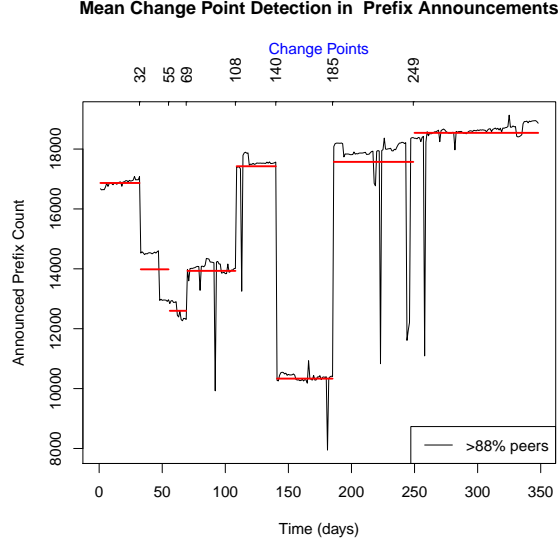}
 \caption{Mean Change Point Detection on Prefix Count}
 \label{fig:meancpt}
\end{figure}
\indent 
\subsection{Change Point Detection}
The Change Point Detection is a clustering mechanism on non stationary time series.
It is the name given to the problem of estimating the point at which the statistical properties of 
a sequence of observation change. In this, a cluster is assumed
to have same mean value. The clusters are identified using Segmentation Neighborhood (SegNeigh) \citep{segneigh} 
algorithm based on mean change with respect to global mean. The maximum number of segments needs to be specified for
 the algorithm.
The Cumulative Sum (CUSUM) is computed on the time series data over which SegNeigh algorithm is applied to
 detect the
mean changed segments. The segments that have mean prefix reachability count varying more than 15 $\%$ is
 considered as significant long term event. This threshold value is subjected to the requirement of 
the Internet Service Providers (ISPs).
 The change point package \citep{R:changepoint} in
 R statistical software is used
to detect change points. The  long and short duration segments are identified with their peak impact values
and given in table \ref{table:cptLS}.
 \begin{table*}
\caption{Mean Change Points and Prefix Reachability Events}
\label{table:cptLS}       % Give a unique label
\begin{center}

\begin{tabular}{|c|c|c|c|c|c|c|}
\hline\noalign{\smallskip}
 Time & Segment & $>$ 88 $\%$  Peer  & Diff  & Long Term & Stochastic & Peak \\
 Period & mean &mean & Prefix  $\%$ & Event& Events & Impact $\%$\\
\noalign{\smallskip}\hline\noalign{\smallskip}
\hline

01/01 - 05/02  & 17407 &  16867 & 97 & No  & 0 &  0 \\ 
10/02 - 03/03 & 17699 &  13983 & 79 & Yes & 0 &  0 \\  
04/03 - 21/03 & 17645 &  12597 & 71 & Yes & 0 &  0 \\  
22/03 - 04/05 & 17648 &  13932 & 75 & Yes & 1 & 56  \\
05/05 - 06/06 & 17864 &  17425 & 97 & No & 1 & 74 \\
07/06 - 21/07 & 18136 & 10333 &  56 & Yes  & 1 & 44 \\
22/06 - 23/09 & 18374 & 17571 &  95 & No  &  2&  59 ,63  \\
24/09 - 31/12 & 18972 & 18540 &  97 & No  & 1  & 58   \\
\noalign{\smallskip}\hline
 \end{tabular}
\end{center}
\end{table*}
\subsection{Temporal Aberration Algorithm}
\begin{figure}
 \centering
 \includegraphics[width=0.5\textwidth]{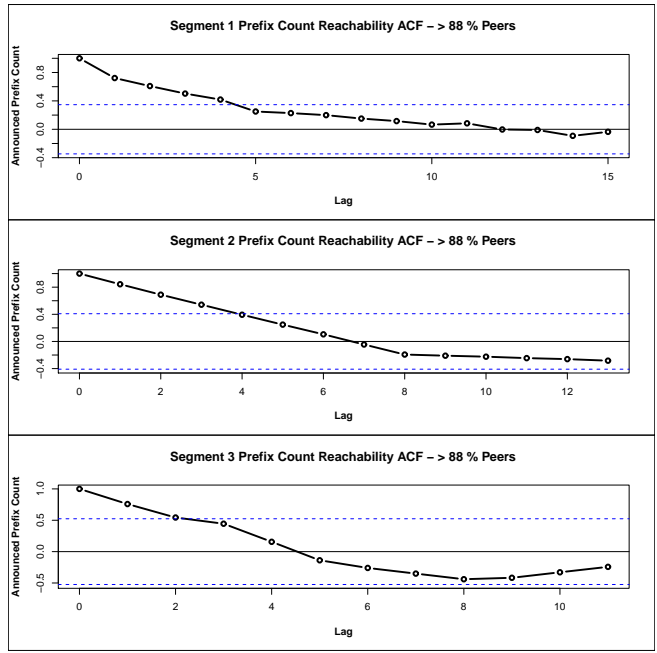}
 \caption{ACF for Announced Prefix Counts of Different Segments}
 \label{fig:segacf}
\end{figure}

\begin{figure}
 \centering
 \includegraphics[width=0.5\textwidth]{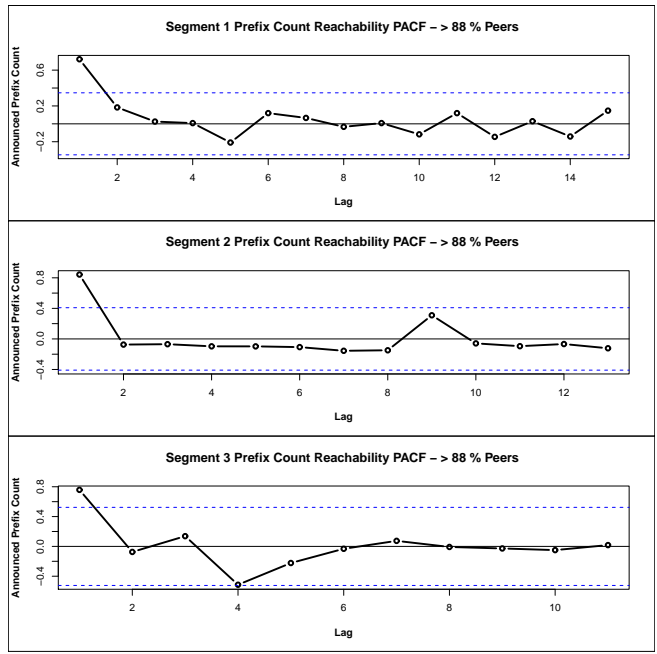}
 \caption{PACF for Announced Prefix Counts of Different Segments}
 \label{fig:segpacf}
\end{figure}

\begin{figure}
 \centering
 \includegraphics[width=0.5\textwidth]{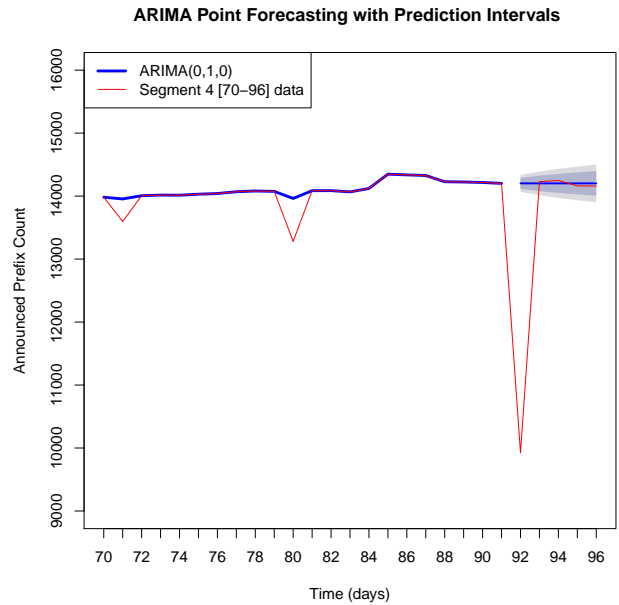}
 \caption{ARIMA Model Forecasting with Prediction Intervals}
 \label{fig:p88_seg4_f}
\end{figure}

The temporal aberrancy detection algorithm is described as algorithm that sequentially evaluate 
the departure of the observed rate of a measure from what would be expected based on previous history.
These algorithms are in general classified based on the underlying methods such as control charts,
regression models, time-series methods and scan statistics \citep{temporalAberration1}.
 Considering our non stationary time series measure, we use ARIMA model based algorithm.\\
\indent  
The algorithm computes the difference between the baseline measure and the current observation.
An event is identified when the difference is greater than a positive threshold or lesser than a negative
threshold. The threshold values are taken from the 95 $\%$ confidence level prediction intervals of the
considered model. ARIMA models have three components namely Auto Regression (AR), Moving Average (MV) and Trend.
This model takes care of the temporal evolution of the prefix announcements.
\\
\indent
By analyzing Auto Correlation Function (ACF) and Partial Auto Correlation Function (PACF) coefficients of each segment,
 the parameter values for AR, MV and Trend components are identified. The ACF and PACF plots for first three segments
are given in figures \ref{fig:segacf} and \ref{fig:segpacf}.   
The lag value that has significant ACF coefficients are assigned for MV parameter of the segment model. For example,
segment 1 has significant ACF coefficient upto lag 4 and hence MV(4) is considered. Similarly the lag value
 that has significant PACF coefficients are assigned to AR parameter of the segment model. In our case none of the
segments have significant PACF coefficients and hence AR(0) is assigned for all the segment models. The trend is linear 
and hence the additive trend is considered in the model. The MV parameter provides the indication on how many past
values are taken to compute the base line. The model forecasts one step ahead and provides prediction intervals used as
thresholds.\\ 
\indent
For brevity, we provide model selection and forecasting for segment 4 which spans from time unit 70 to 108 in the
 time series.
The ACF coefficient of this segment is nonsignificant in all the lags except 0. Similarly PACF coefficient
is also non significant in all the lags.
So we use a special case of ARIMA (random walk model) which is represented by ARIMA(0,1,0). 
We fit the data to ARIMA(0,1,0) from the time unit 70 to 91 and show the five step ahead forecast for
 time units 92 to 96 for illustration. The segment data in time unit 92 is 
 stochastic drop event since it deviates significantly from 95 $\%$ prediction interval.
 The model forecasting and prediction intervals are given in figure \ref{fig:p88_seg4_f}.
The point forecasting and prediction intervals for  80 and 95 $\%$ confidence levels are given in table \ref{table:p88_seg4_f}.
 \begin{table*}
\caption{ARIMA(0,1,0) Point Forecasting and Prediction Intervals}
\label{table:p88_seg4_f}       % Give a unique label
\begin{center}

\begin{tabular}{|c|c|c|c|c|c|c|c|}
\hline\noalign{\smallskip}
 Time (days) & Actual & Point  & 80  $\%$ Low &  80  $\%$ High &  95  $\%$ Low & 95  $\%$ High  \\
 \noalign{\smallskip}\hline\noalign{\smallskip}
\hline
92 & 9925    &14202 & 14114.64 & 14289.36 & 14068.39 & 14335.61 \\
93 & 14229   &14202 & 14078.45 & 14325.55 & 14013.04 & 14390.96 \\
94 & 14248   &14202 & 14050.68 & 14353.32 & 13970.58 & 14433.42 \\
95 & 14163   &14202 & 14027.27 & 14376.73 & 13934.78 & 14469.22 \\
96 & 14160   &14202 & 14006.65 & 14397.35 & 13903.23 & 14500.77 \\

\noalign{\smallskip}\hline
 \end{tabular}
\end{center}
\end{table*}

From the figure \ref{fig:p88_seg4_f} and the table \ref{table:p88_seg4_f}, it can be inferred that one step ahead 
forecast can clearly capture the deviant behaviors. Except the data point at 92, the data spans from 93 to 96 are
well within the 95 $\%$ prediction intervals.
The forecast package \citep{R:changepoint,R:forecast} in
 R statistical software is used to do one step ahead forecast for the ARIMA model. By iteratively calling the
forecast function of the model, stochastic events are identified and quantified.
The identified short term events and their impact are given in table \ref{table:cptLS}.
\section{Validation}
\label{sec:5}
\begin{figure}
 \centering
 \includegraphics[width=0.5\textwidth]{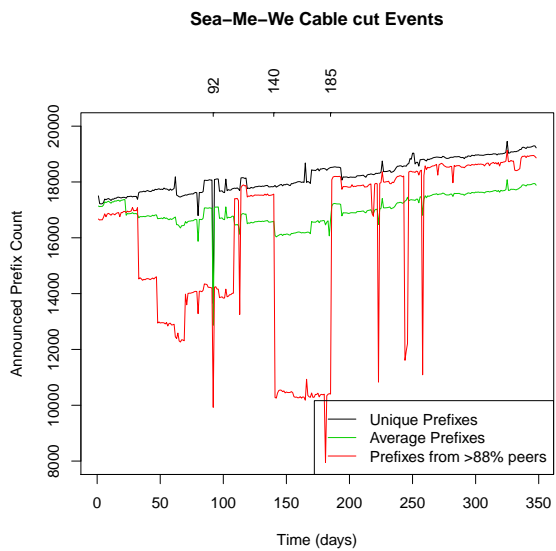}
 \caption{Announced Prefix Counts from Different Measures}
 \label{fig:comparePrefix}
\end{figure}

Using the change point detection technique, we  have identified 4 long term events. Legitimate
route change events due to link status change usually affect large number of prefixes \citep{prefixhijack1} while
route changes due to prefix hijacking usually target specific network prefixes. 
In our case, each of the detected events
has more than 15 $\%$ prefix reachability reduction and prevails for more than 10 days. Hence the events should be
 related with link status change induced by cable cuts. 
 The time period associated with the  fourth  event (segment 6) is correlated with SEA-Me-We4 cable cut and restoration
duration from 07/06/2012 to 21/07/2012 \citep{SMW42012Dec2}. The total prefix paths to Indian ASes
computed from route-view table also indicates a significant path reduction during this period. 
 The stochastic event identified in segment 4 is matching
with the same cable cut event occurred on 17/04/2012 that got restored within a day.\\ 
\indent  
We have computed the average number of prefixes announced to Indian AS topology by dividing the total prefix 
paths using average number of peers.
The uniquely filtered prefixes, average number of prefixes and prefixes announced by $>$ 88 $\%$ of peers are given
in figure \ref{fig:comparePrefix}. 
All the three measures have the indication for the cable cut events, but with different levels of noise. 
The route change events  can also be observable in the Upstream Betweenness Centrality (UBC) change to an AS under
 consideration. Here betweenness centrality is computed as number of paths passing through an upstream 
to the prefixes of particular AS. In the cable cut event, it is reported that the Indian transit AS 9498
has severe impact on its prefix reachability. Hence we computed and  analyzed the UBC measure for AS 9498 
from the extracted AS paths of Indian AS topology. A maximum of 25 upstreams (some of them may be peers)
 are found through which the prefixes of the AS can be reachable from global locations. The first 4 upstreams based
on the decreasing order of the UBC value is given in table \ref{table:9498reachability}. In this $UBC_{i}$ represents
the paths passing through $i_{th}$ upstream. On 17th April 2012 there is
a decrease in the UBC measure for all the upstreams with approximate path drop  of 3500. No change in upstream count 
and the ordering of the upstream is observed. But on 7-06-2012 we observed a change in the upstream order which is 
based on the UBC value and also a peer drop. In the top three upstreams, there is a path increase to the magnitude
of 2500 and in middle few upstreams there is a path drop to the magnitude of 1000. The dropped peer UBC is 
approximately 5500. Over all there is a path drop of around 4000 is observed to AS 9498. Further more, path drops 
fluctuating
 around this value is found upto 21-7-2012. There may be other factors involved in change of paths during this period, 
but the upstream order change and peer drop evidences show that the cable cut has significant impact on the 
reachability of the prefixes during this period.
\begin{table*}
\caption{Prefix Reachability to AS9498 Via Different Upstreams}
\label{table:9498reachability}       % Give a unique label
\begin{center}

\begin{tabular}{|c|c|c|c|c|c|c|c|c|}
\hline\noalign{\smallskip}
 Date & $Upstream_{1}$  & $Upstream_{2}$ & $Upstream_{3}$  & $Upstream_{4}$ & $UBC_{1}$ & $UBC_{2}$ & $UBC_{3}$ & $UBC_{4}$  \\
 \noalign{\smallskip}\hline\noalign{\smallskip}
\hline
16-04-2012&  3561 &  174 & 3549 &  3257 &  21134 &  20527 & 17678 &  16570 \\ 
17-04-2012&  3561 &  174 & 3549 &  3257 &  17260 & 16603 & 14324 &  13204  \\
18-04-2012&  3561 &  174 & 3549 &  3257 &  21154 &  20628 & 17571 &  16900 \\ 
05-06-2012&  3549 & 3561 & 174 &  3257 &  19715  & 17981 & 16272 &  15990  \\ 
06-06-2012&  3549 & 3561 & 174 & 3257 &  19954 & 18705 & 16455 &  16169  \\  
07-06-2012&  3561 & 3549 & 174 &  3257 &  21128 &  19889 & 16620 & 16119\\  
08-06-2012&  3561 & 3549 & 174 &  3257 &  20250  & 20111 & 16478 & 16210 \\ 
\noalign{\smallskip}\hline
 \end{tabular}
\end{center}
\end{table*}

\section{Conclusion}
\label{sec:6}
In this work, we have analyzed the route-view data for distribution of peer locations from which BGP feeds are
collected to the route-view table. The prefix announcements from different spatially located peers are interpreted as
 reachability of prefixes from those locations. The peer distribution analysis on different continents
reveals that the number of representative peers in Asia, Australia and Africa is less than 10 $\%$ of the 
total peers involved in route collection. Hence to understand the events occurring in these 
regions, we require BGP feeds from more representative peers in the regions. Prefix announcements for Indian AS topology
 is extracted 
from the daily route-view snapshots for a duration of 1 year. The temporal prefix reachability patterns to Indian ASes
from each of the peers are analyzed. The temporal patterns for 20 peers out of 38 are found to be similar
 with respect to 
 direction and magnitude of the events.  The patterns from other 18 peers exhibit different number of prefix raise
 and drop
events. To capture the events occur spatially in a single temporal pattern and quantify the impact and duration, 
 we proposed a counting based
spatio temporal measure on prefix reachability from \textbf{\emph{x}} $\%$ of peers.  
Using this measure we detect long term events by employing change point detection technique and quantify the
 impact of each event. The stochastic events are captured using temporal aberration detection algorithm
that use ARIMA models on the segment under consideration. The sensitivity and specificity of the mechanisms
 are validated using the  Sea-Me-We4 cable cut events reported in the news blogs. Our future work is to identify 
the root causes for the other events that induced the aberration in the measure and verify the performance of the 
mechanisms over the measure with respect to India as well as other countries in the APNIC region.

%\begin{acknowledgements}
%\end{acknowledgements}

\bibliographystyle{spbasic}      % basic style, author-year citations
\bibliography{controlplane_events} % name your BibTeX data base

%\begin{thebibliography}{}
% etc
%\end{thebibliography}

\end{document}